\documentclass[pra,twocolumn,showpacs,floatfix,superscriptaddress]{revtex4-1}
\usepackage[utf8]{inputenc}
\usepackage[T1]{fontenc}
\usepackage{ae}
\usepackage{amsmath,amssymb,amsfonts}
\usepackage{mathrsfs}
\usepackage{units}
\usepackage[squaren]{SIunits}
\usepackage{bm,bbm}
\usepackage{bbm}
\usepackage{units}
\usepackage{graphicx}
\usepackage{dcolumn}
\usepackage{float}
\usepackage{mathrsfs}
\usepackage[cmyk]{xcolor}

\renewcommand{\vec}[1]{\bm{#1}}							

\begin{document}

\title{Mean-field predictions for a dipolar Bose-Einstein condensate with $^{164}$Dy}

\author{Damir Zajec}
\email{zajec@itp1.uni-stuttgart.de}
\author{G\"unter Wunner} 
\affiliation{Institut f\"ur Theoretische Physik 1, Universit\"at Stuttgart,
70550 Stuttgart, Germany}

\date{\today}

\begin{abstract}

Dipolar Bose-Einstein condensates are 
systems well-suited for the investigation of effects caused by 
the non-local and anisotropic dipole-dipole interaction. 
In this paper we are interested in properties which
are directly connected to the realization of a condensate 
with $^{164}$Dy, such as stability and phase diagrams. 
Additionaly, we study the expansion of dipolar condensates, 
and find signatures of the dipole-dipole interaction in terms
of structured states and a deviation of the well-known 
inversion of the aspect ratio of the cloud during a 
time of flight. Our analysis is based on the extended
Gross-Pitaevskii equation, which we solve numerically 
exact on a grid by means of an imaginary- and real-time evolution.

\end{abstract}
\maketitle

\section{Introduction}
\label{sec:introduction}

The realization of dipolar Bose-Einstein condensates (BECs)
with large magnetic moments is an important step towards 
the opportunity for a more detailed study of the effects 
caused by the nonlinear, anisotropic, non-local, and long-ranged
dipole-dipole interaction (DDI), such as structured ground 
states \cite{Ronen2007,Koeberle2009}, solitons 
\cite{Tikhonenkov2008,Pedri2005,Koeberle2012,Eichler2012}, and 
roton-maxon spectra \cite{Santos2003}. Dipolar condensates 
have recently been realized with atoms of $^{52}$Cr 
\cite{Griesmaier2005,Beaufils2008,Lahaye2009}, 
$^{164}$Dy \cite{Lu2010,Lu2011} and $^{168}$Er \cite{Aikawa2012}. 
Furthermore, there has been vast progress in the realization 
of a condensate with polar molecules with an 
electric dipole moment \cite{Ni2008}. 

Two examples for the recent study of effects due to the DDI
have been the experimental investigation of the collapse dynamics
\cite{Lahaye2008} and the inversion of the aspect ratio of 
the cloud \cite{Lahaye_2007} of a dipolar $^{52}$Cr BEC. 
In the former case the scattering length was ramped down 
below the collapse threshold and absorption images were 
made after the collapse and explosion of the condensate. 
The density distribution inherited the d-wave symmetry 
$1-3 \cos^2{\vartheta}$ of the DDI. In the latter case 
the dipolar forces were shown to inhibit the inversion of 
the aspect ratio of the cloud, which usually occurs 
during a time of flight. An exhaustive review of the 
physics of dipolar quantum gases has recently been given 
by Lahaye \textit{et al.} \cite{Lahaye2009} and 
Baranov \textit{et al.} \cite{Baranov_2012}. 

In this paper, we use the extended Gross-Pitaevskii equation 
to calculate the dynamics of a condensate, which yields
an accurate description for sufficiently low temperatures and reads 
    
\begin{align}
  \label{eq:GPE}
  H \psi(\vec{r},t) &= \left( -\frac{1}{2}\Delta + V_{\mathrm{har}} +
    V_{\mathrm{dd}} + V_{\mathrm{sc}} \right) \psi (\bm{r},t) \nonumber \\
  &= \mathrm{i} \partial _{t}\psi (\bm{r},t)\,,
\end{align}
with
\begin{align}
  V_{\mathrm{har}} &= \frac{1}{2}(\omega_x^2 x^2 + \omega_y^2 y^2 + 
  \omega_z^2 z^2) \,,\nonumber \\
  V_{\mathrm{dd}} &= 3Na_{\mathrm{dd}} \int \text{d}^3r'
  \frac{1-3\cos^2\vartheta}{|\vec{r}-\vec{r}'|^3}
  |\psi(\vec{r}',t)|^2 \,,\nonumber \\
  V_{\mathrm{sc}} &= 4\pi Na |\psi(\vec{r},t)|^2 \,.\nonumber
\end{align}
Here $N$ is the number of particles, $a_{\mathrm{dd}}$ is the 
dipole length, and $a$ denotes the scattering length. 
The dipoles are aligned along the $z$-axis, such that 
$\vartheta$ is the angle between the $z$-axis and the 
vector $\vec{r}-\vec{r}'$. This form implies that all 
lengths are scaled in units of $\sqrt{\hbar/m}$ and all 
energies in units of $\hbar$. The quantities $Na$ 
and $Na_{\mathrm{dd}}$ control the strength of the scattering 
and the dipole interaction, respectively, and are assumed 
to be independently adjustable. The ratio $\epsilon=a_{\mathrm{dd}}/a$ 
is a measure for the importance of the dipole interaction. 
The strength of short-ranged interactions, which can be described 
by a Fermi-like contact interaction, can be controlled by means 
of a Feshbach-resonance \cite{Chin2010,Inouye1998}. An external 
magnetic field changes the coupling between a bound state and a 
state of an incoming particle and allows for the control of 
the scattering length. This can be used to increase 
$\epsilon$ and therefore to enhance the dipolar character 
of the condensate. However, since Feshbach-resonances reduce 
the lifetime of a condensate, one prefers the realization of 
condensates with larger magnetic moments. Of all the above 
mentioned species with permanent magnetic dipole moments, 
$^{164}$Dy stands out with $\mu=10\mu_{\mathrm{B}}$, 
where $\mu_{\mathrm{B}}$ is the Bohr magneton.

Our results can be scaled and are therefore valid for all 
dipolar systems. However, we are in particular interested in 
a condensate with $^{164}$Dy, which means that in the 
following all parameters are adjusted to fit the needs 
of a condensate with this species. To the best of 
our knowledge, this is the first fully numerical three-dimensional study 
for a realistic set of parameters close to those in actual 
experimental setups. We emphasize that our calculations do not 
depend on any further simplifications or restrictions beyond the 
use of a mean-field ansatz.

In the next section we give a short introduction to our 
numerical method and present our results with respect to 
expansion dynamics, structured ground states and a 
self-induced Josephson junction.

\section{Condensate wave functions}

We solve the Gross-Pitaevskii equation numerically exact 
on a grid by means of the split-operator method
using the imaginary- and real-time evolution. 
This yields a series of Fourier transforms, which we evaluate 
with a highly parallelized algorithm using the 
CUDA-architecture of NVIDIA. A more detailed overview is 
given in \cite{Eichler2012}. For the calculation of stability 
and phase diagrams we start with a simple Gaussian for a set 
of parameters far away from the stability border. 
This border is given by the maximum value (in this paper either 
the angle of rotation $\alpha$ or the number of particles $N$) 
for which the imaginary-time evolution converges and a stationary 
ground state exists. After the calculation of the ground state, 
we use this wave function as the initial wave function 
for the calculation of the next set of parameters, where we keep 
the aspect ratio or $\epsilon$ constant until we reach the 
stability border. Since our computational ressources allow 
for a very small step size, we can make very precise predictions 
with respect to stability and phase diagrams.

\subsection{Expansion dynamics}

\begin{figure}[tbp]
 \includegraphics[trim= 0cm 0cm 0cm 0cm, clip=true, width=0.9\columnwidth]{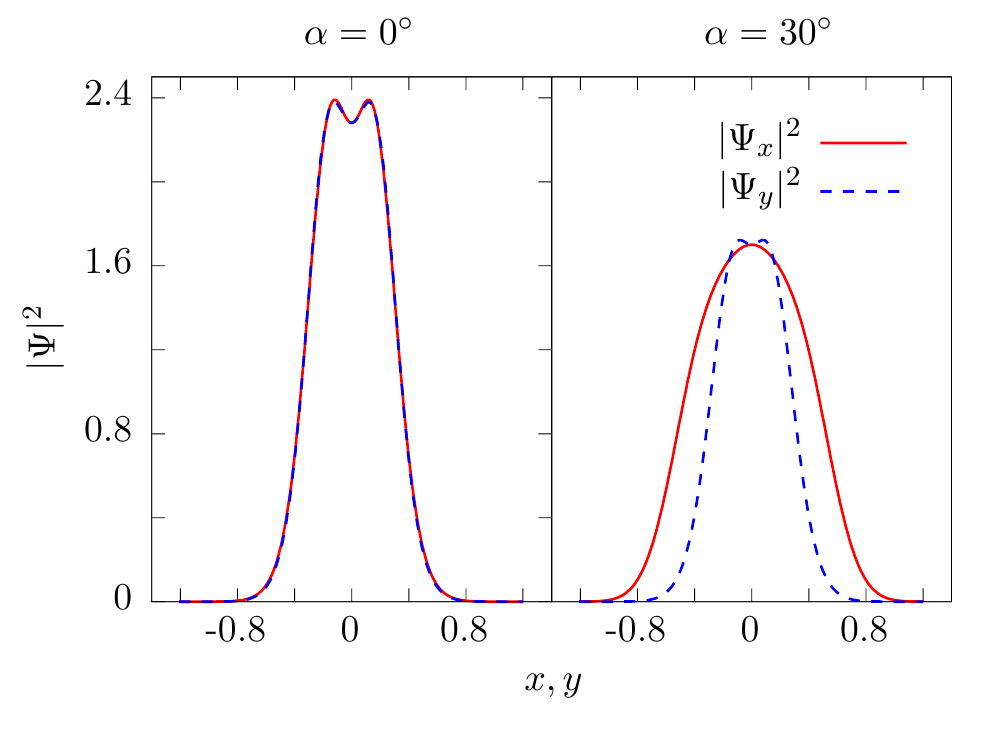}
 \caption{(Color online) One-dimensional absorption images 
           in $z$-direction for $\alpha=0\degree$ and 
           $\alpha=30\degree$ after 6 ms of TOF. 
           Note that for $\alpha = 0\degree$, the graphs for the 
           quantities $|\Psi_x|^2$ and $|\Psi_y|^2$ are congruent.
	   Since the visibility $c$ of the structures 
           decreases with an increase of $\alpha$ the largest 
           value of $c=0.023$ is given for $\alpha=0\degree$.
	   This is below the benchmark of $c \approx 0.1$, 
           which is a necessary condition for the experimental 
           observation of such structures. Additionaly, we see 
           that the rotation of the trap destroys the 
           rotational symmetry in the $xy$-plane.
  }
\label{fig:1D_absorption_image}
\end{figure}

One possible setup for the experimental investigation of dipolar 
signatures in condensates is to tilt the polarization axis 
with respect to the external confinement by means of an 
external magnetic field and investigate its expansion dynamics. 
The influence of the DDI should manifest itself in a deviation 
of the expansion dynamics as compared to a condensate which is 
strongly dominated by isotropic short-ranged interactions. The 
most clear evidence would be a TOF (time of flight) sequence in 
which the direction of the largest width of the cloud
coincides with the polarization axis, regardless of the initial 
confinement. In this paper we study the equivalent scenario, where 
we rotate the external trap and the dipoles remain permanently 
aligned in $z$-direction. A tilt of the polarization axis clockwise 
in the $xz$-plane corresponds to a rotation of the trap anti-clockwise 
along the $y$-direction. The corresponding 
transformation of the coordinates reads

\begin{align}
 x' &= x\ \mathrm{cos}\ \alpha + z\ \mathrm{sin}\ \alpha, \nonumber \\
 y' &= y, \\
 z' &= -x\ \mathrm{sin}\ \alpha + z\ \mathrm{cos}\ \alpha. \nonumber
\end{align}
\begin{figure}[tb]
 \includegraphics[trim= 1.5cm 8cm 2cm 4cm, clip=true, width=1.0\columnwidth]{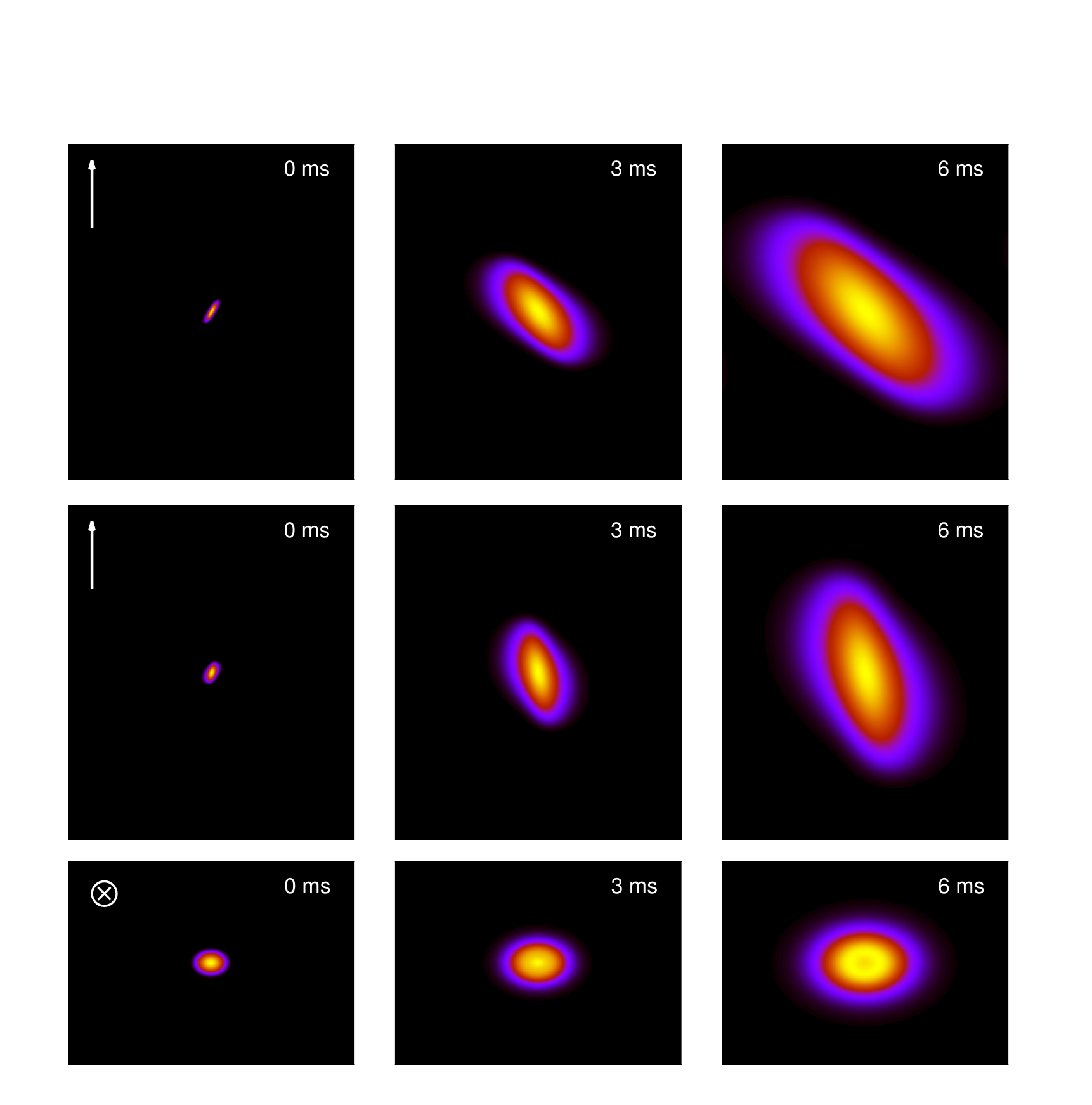}
 \caption{(Color online) Absorption images in $y$- and 
   $z$-direction ($|\psi|^2$ integrated
   along the $y$- and $z$-axis, respectively) for three different 
   expansions of the condensate at times $t=0, 3$ and 6 ms 
   (from left to right). The symbol in the left top corner illustrates 
   the polarization axis, which always points into the 
   $z$-direction. Top: Absorption images in $y$-direction 
   ($x$ abscissa, $z$ ordinate) $\bot$ polarization axis with 
   $\omega_{x,y,z} = 2\pi \cdot (205,195,760)$ Hz, 
   $\epsilon=1.1$ and $\alpha = 60\degree$. The expansion is 
   dominated by the anisotropy of the momentum distribution.
   Center: Absorption images in $y$-direction
   ($x$ abscissa, $z$ ordinate) $\bot$ polarization axis with 
   $\omega_{x,y,z} = 2\pi \cdot (205,195,360)$ Hz, $\epsilon=1.1$ 
   and $\alpha = 45\degree$. Here the anisotropy in the momentum 
   distribution is lower than in the upper case, which yields an 
   expansion dynamics showing a clear signature of the 
   dipole-dipole interaction. Bottom: Absorption images in 
   $z$-direction ($x$ abscissa, $y$ ordinate) \textbardbl \ polarization axis
   with $\omega_{x,y,z} = 2\pi \cdot (205,195,760)$ Hz, 
   $\epsilon=1.1$ and $\alpha = 0\degree$. Here the occurence of 
   structured states during the TOF is recognizable 
  (see also Fig.~\ref{fig:1D_absorption_image}). 
  }
\label{fig:expansion}
\end{figure} 
Fig.~\ref{fig:expansion} shows absorption 
images of three TOF sequences with $\epsilon=1.1$ and different 
rotation angles $\alpha$. For absorption images in the $y$- and 
$z$-direction we integrate $|\psi|^2$ along the $y$- and $z$-axis, 
respectively. The first seqence at the top shows a TOF for initial 
harmonic trapping with $\omega_{x,y,z} = 2\pi \cdot (205,195,760)$ Hz 
and $\alpha=60\degree$, where we have used the trapping parameters 
from \cite{Lu2011}. Since an anisotropic confinement yields an 
anisotropic momentum distribution, we see an inversion of the aspect 
ratio of the cloud. The TOF does not show any signature of the 
dipole-dipole interaction since the dynamics of the condensate 
is dominated almost entirely by the anisotropic momentum distribution. 
Therefore in the middle panels we reduce the aspect ratio of the 
trap and adjust the confinement to 
$\omega_{x,y,z} = 2\pi \cdot (205,195,360)$ Hz and $\alpha=45\degree$.
For this setup the dynamics of the condensate is in stark contrast 
to the TOF above since the expansion seems to be governed by the 
direction of the strongest initial confinement and the polarization 
axis. This can be cleary seen by the fact, that the direction of 
smallest width for 0 ms is not the direction which dominates the 
TOF, as is the case for the sequence above. Therefore this 
expansion dynamics strikes a balance between the direction 
that is given by the anisotropic momentum distribution and the 
polarization direction which is energetically preferable. 
The last TOF shows absorption images in $z$-direction for 
$\omega_{x,y,z} = 2\pi \cdot (205,195,760)$ Hz and $\alpha=0\degree$. 
Here we see a blood-cell structured condensate where the density 
peak is away from the center \cite{Koeberle2009}, which occurs 
about 6 ms after the release from the trap (for better visibility 
see Fig.~\ref{fig:1D_absorption_image}). Note that the initial 
density distribution in the $xy$-plane is given by a Gaussian. 
Therefore the structured state is not an artifact from the 
stationary ground state but rather a product of the expansion dynamics.

The crucial parameter for the experimental oberservation 
of structured states is the visibility 

\begin{equation}
c = \frac{n_{\text{max}} - n_{\text{min}}}{n_{\text{max}} + n_{\text{min}}},
\label{visibility}
\end{equation}
where

\begin{equation}
n=\int \text{d}z |\psi(\bm{r})|^2,
\end{equation} 
with $n_\text{max}$ as the maximum value and $n_\text{min}$ as the 
value in the center of the plane. The visibility $c$ should be at 
least at about 0.1 to allow for a high-contrast imaging of 
the density distribution. It turns out that for all angles 
$\alpha$ the visibility is smaller than this benchmark. 
For the smallest angle $\alpha=0\degree$ we find $c=0.023$, 
whereas for $\alpha=30\degree$ the corresponding visibility 
is $c=0.006$. Additionaly, as can be seen in 
Fig.~\ref{fig:1D_absorption_image}, where we depict the quantities 

\begin{align}
  |\Psi_{x}|^2 & = \int \text{d}z |\psi(x,0,z)|^2 \nonumber \\
  |\Psi_{y}|^2 & = \int \text{d}z |\psi(0,y,z)|^2, 
  \label{psi_square}
\end{align}
the rotation of the external confinement destroys the rotational 
symmetry in the $xy$-plane and leads to a saddle-shaped structured state.   

The small value of $c$ during the expansion dynamics 
suggests that such structured states can only be seen 
if modern imaging systems are further improved or
systems with very large dipolar moments such as polar 
molecules are used.

\subsection{Structured ground states}

\begin{figure}[tbp!]
 \includegraphics[trim= 0cm 0cm 0cm 0cm, clip=true, width=1.0\columnwidth]{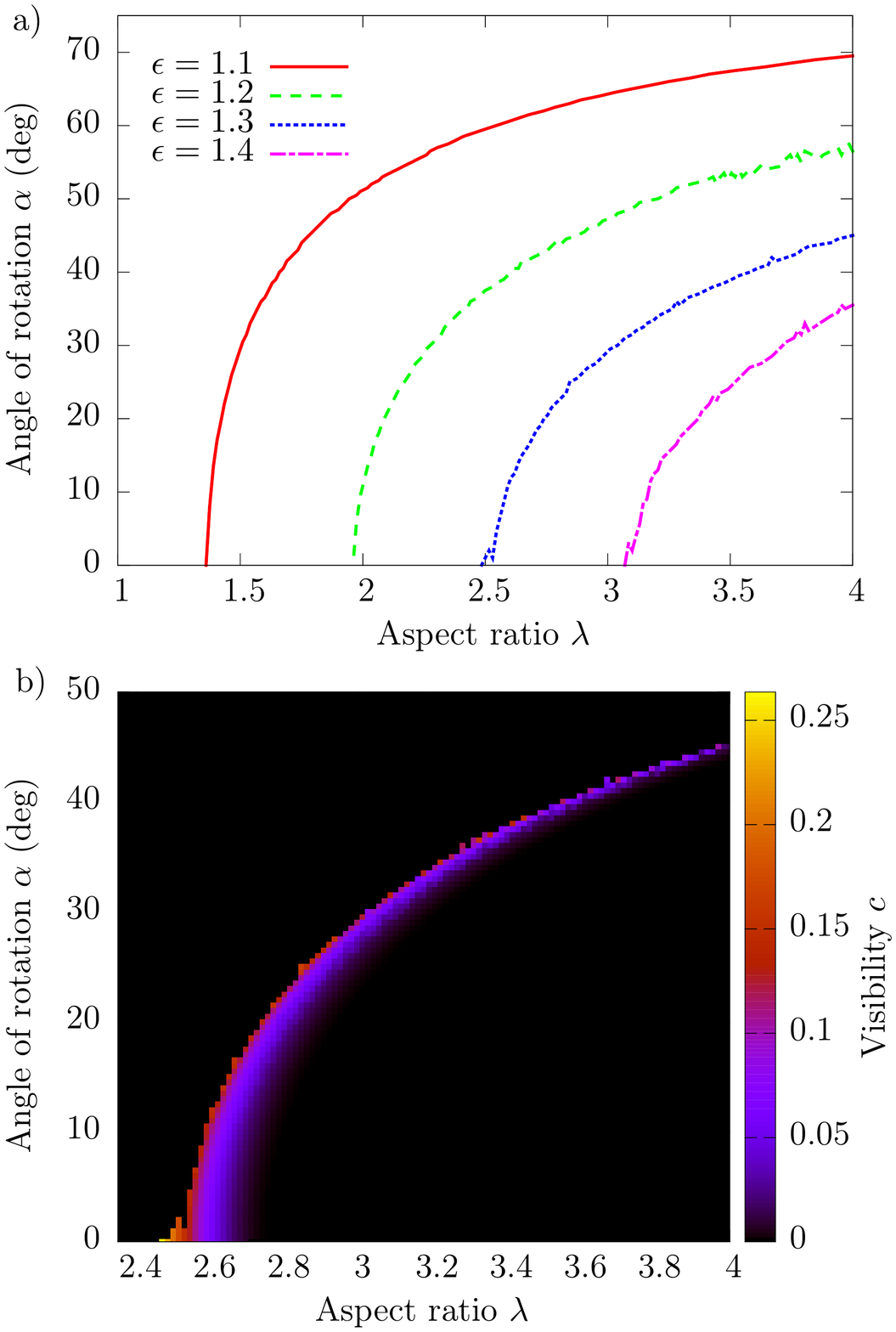}
 \caption{(Color online) a) Stability diagrams for four values 
           of $\epsilon$ with respect to the angle of rotation 
           $\alpha$ and the aspect ratio of the trap 
	   $\lambda = \omega_z / \omega_{x,y}$ with $\omega_z= 2\pi \cdot 760$ Hz, 
           and $Na_\mathrm{dd}=3.6$ (which corresponds to $10^4$ particles). 
           We have plotted the maximum value of $\alpha$ which 
           allows for a stable condensate for a given aspect 
           ratio $\lambda$. A smaller value of $\epsilon$ stabilizes 
           the condensate against the unstable head-to-tail 
           configuration, which corresponds to large angles of 
           rotation and small aspect ratios. b) Phase diagram 
           for the visibility $c$ for $\epsilon = 1.3$. Structured 
           stationary ground states appear at the border 
           of stability. Since we find $c \approx 0.1$ for a 
           broad strip, an experimental observation should be possible.
  }
\label{fig:stability_visibility}
\end{figure} 

Structured states are not only a feature of dynamical processes, 
but do occur as stationary ground states as well. 
In this section we investigate where in the parameter 
space these structured states occur and if their 
visibility is large enough to satisfy 
the above mentioned criterion of $c \gtrsim 0.1$. 
Fig.~\ref{fig:stability_visibility} depicts the 
stability border for four different values of 
$\epsilon$ and a phase diagram of the visibility of 
structured ground states for $\epsilon=1.3$ with respect 
to the angle of rotation $\alpha$ and the aspect ratio of 
the trap $\lambda = \omega_z / \omega_{x,y}$ with 
$\omega_z= 2\pi \cdot 760$ Hz, and $Na_\mathrm{dd}=3.6$, 
which corresponds to $10^4$ particles. An increase of the 
angle of rotation or a reduction of the aspect ratio 
results in more dipoles in the attractive head-to-tail 
configuration, and thus destabilizes the condensate. 
These parameters are therefore crucial for the stability 
properties of a dipolar condensate and can be used to 
investigate its stability borders. 

Experimentally one would prefer to change the axis of 
polarization as opposed to the aspect ratio of the trap 
since this axis can be easily controlled by an external 
magnetic field. As one would expect, smaller values of 
$\epsilon$, which correspond to a condensate less dominated 
by the DDI, stabilize the condensate against larger angles 
of rotation and smaller aspect ratios. 

Structured states can be found in the vicinity of the border 
of stability. Here we find a broad strip in which 
$c \approx 0.1$. For low values of $\alpha$, these
structures resemble the blood-cell-like structured condensates
(in the plane $\bot$ polarization axis) which also occur during the 
TOF (see Fig.~\ref{fig:expansion}). The spatial size of the 
structures is limited by the external confinement. Here the diameter of the 
ring $\gtrsim$ 1 $\mu$m. This should be sufficient for an experimental observation.  
    
\begin{figure}[tbp]
 \includegraphics[trim= 0cm 0cm 0cm 0cm, clip=true, width=1.0\columnwidth]{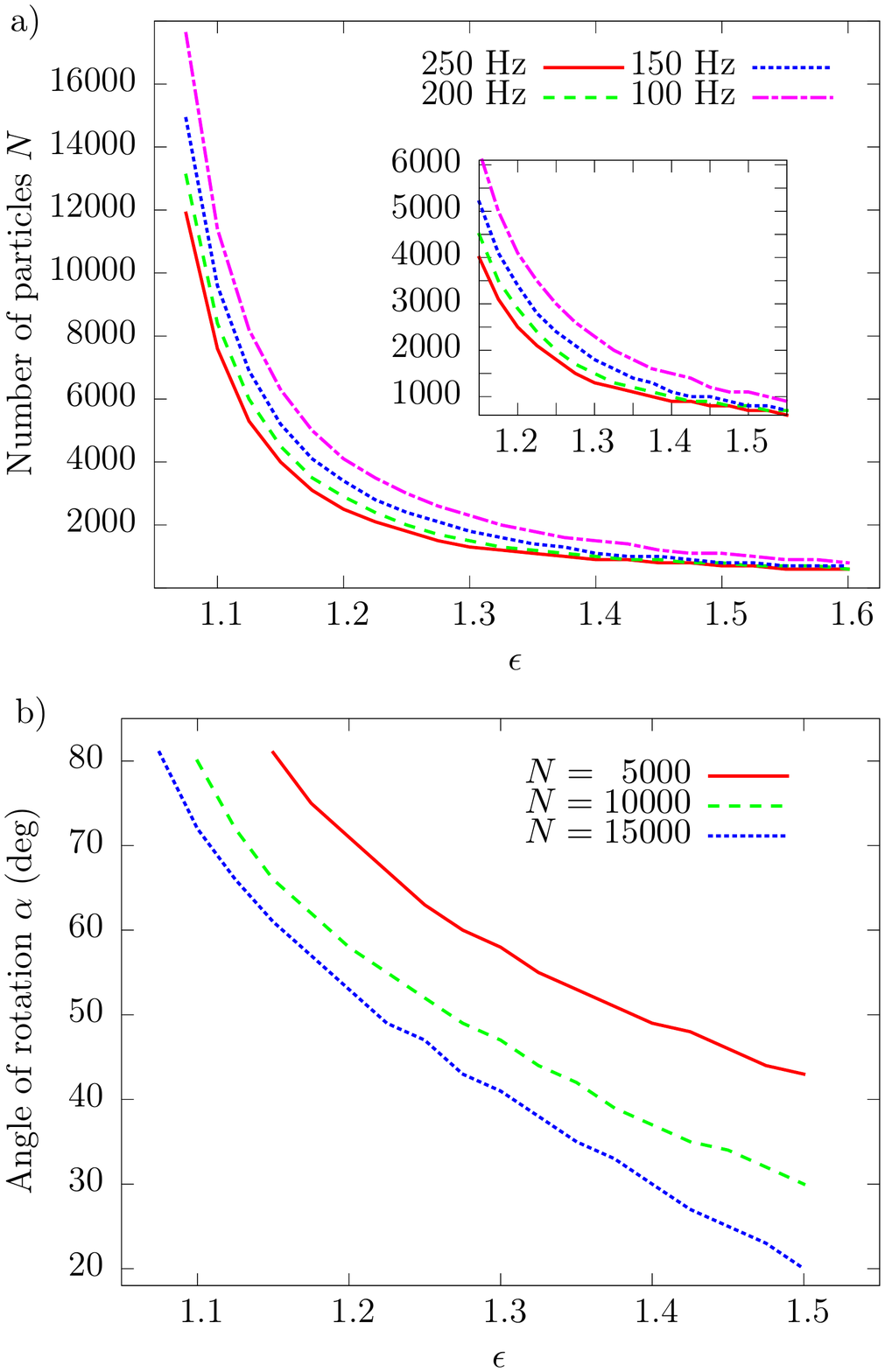}
 \caption{(Color online) a) Stability diagram with respect 
           to the number of particles $N$ and $\epsilon$. 
           Here we have chosen $\alpha=90\degree$, $\omega_{x,y}=2\pi \cdot 50$ Hz 
           and four different values of $\omega_z$, where 
           we have plotted the maximum value of $N$ for 
           a given ratio $\epsilon$ which allows for a 
           stable condensate. An increase of $N$ or $\epsilon$ 
           destabilizes the condensate due to the DDI. b) Stability 
           diagram with respect to the angle of rotation $\alpha$ and 
           $\epsilon$ for $\omega_{x,y,z}=2\pi \cdot (50,50,250)$ Hz 
           and three different values for $N$, where we have plotted 
           the maximum admissible value of $\alpha$. Reducing $\alpha$ 
           stabilizes the condensate und therefore allows for a greater 
	   amount of particles in the condensates.
  }
\label{fig:stability_josephson}
\end{figure} 

\subsection{Self-induced Josephson junction}

One fascinating feature of dipolar condensates is the 
possibility of a self-induced Josephson junction, where 
the effective potential consisting of the external confinement 
and the dipolar potential has the form of a double well. 
These systems show well known phenomena such as 
Josephson oscillations or running phase modes \cite{Abad2011}. 
In this subsection we are again interested in stability borders, 
now with respect to the number of particles $N$, 
the angle of rotation $\alpha$, and $\epsilon$. The self-induced 
Josephson junction can be realized by adding a toroidal potential 
to the external confinement. The external potential then reads

\begin{equation}
  V_{\text{ext}} = \frac{1}{2} (\omega_x^2 x^2 + \omega_y^2 y^2 + 
     \omega_z^2 z^2) + V_0 e^{-2 (x^2+y^2)/\sigma_0^2},
\end{equation}
with $V_0 = 3.286 \cdot 10^5$ (2.5 $\mu$K) and $\sigma_0=0.0764$ (5 $\mu$m). 
For $\alpha=0\degree$ the Gaussian potential is in the 
$xy$-plane and therefore perpendicular to the axis of polarization. 
This is the most stable configuration of the toroidal confinement 
since most of the dipoles are in the repulsive side-by-side configuration 
and arranged in a circle. If we rotate the trap until we attain 
$\alpha=90\degree$, the axis of polarization and the Gaussian 
potential will be in the same plane and we obtain the anisotropic 
density distribution which corresponds to a double-well potential.  

Fig.~\ref{fig:stability_josephson} depicts the stability 
diagrams with respect to the number of particles $N$ and $\epsilon$
as well as the angle of rotation $\alpha$ and $\epsilon$. 
For the first diagram we have chosen $\alpha=90\degree$ and 
$\omega_{x,y} = 2\pi \cdot 50$ Hz and depict the 
stability diagram for four different values of $\omega_z$. 
As can be seen, self-induced Josephson junctions cannot be 
realized with large numbers of $^{164}$Dy particles. 
But it is possible to increase the number of particles by 
reducing the harmonic confinement $\omega_z$. This leads to 
more dipoles in the repulsive side-by-side configuration 
and thus stabilizes the condensate. A more effective 
way of stabilizing the condensate is to choose a smaller 
angle of rotation. For the stability diagram with respect 
to $\alpha$ and $\epsilon$ we have chosen three different 
values for the number of particles and 
$\omega_{x,y,z}=2\pi \cdot (50,50,250)$ Hz. We see again 
that reducing the number of particles stabilizes 
the condensate, and also that the reduction of $\alpha$ 
allows for a greater amount of particles in the condensate.  

\begin{figure}[tbp]
 \includegraphics[trim= 2cm 13cm 5cm 10cm, clip=true, width=1.0\columnwidth]{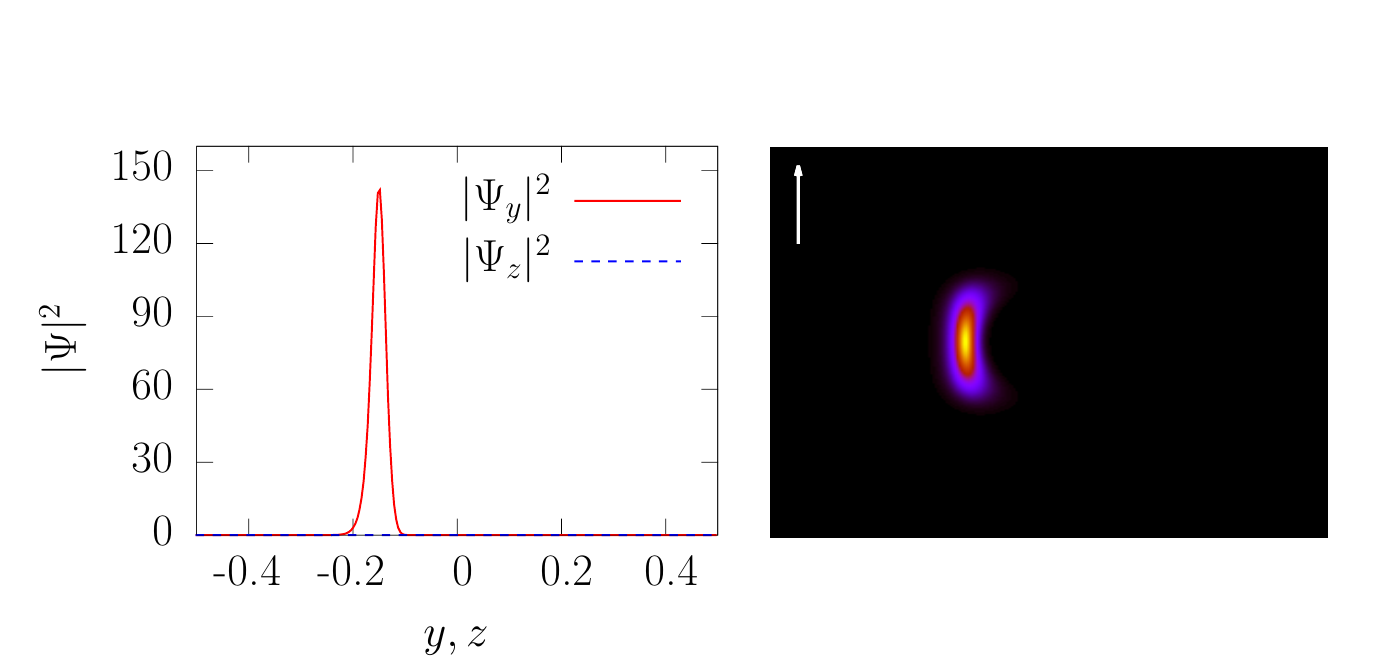}
 \caption{(Color online) One- and two-dimensional absorption 
           image in $x$-direction ($y$ abscissa, $z$ ordinate) 
	   $\bot$ polarization axis with $\alpha=90\degree$, 
           $\omega_{x,y,z}=2\pi \cdot (50,50,250)$ Hz, 
	   $\epsilon=1.5$, and $N=700$. 
	   The symbol in the left top corner illustrates 
           the polarization axis, which points into the $z$-direction. 
           The density distribution has the structure of a macroscopic 
           quantum-self trapping state.
  }
\label{fig:mqst}
\end{figure} 
The calculation of stability diagrams allows for the 
search of new quantum states along the stability border 
of the self-induced Josephson junction. Indeed, we find a 
new stationary state whose density distribution corresponds 
to the macroscopic quantum self-trapping (MQST) state known from 
double-well potentials \cite{Raghavan1999}, in which only 
one well is occupied. We find this state in a small strip along 
the stability border in Fig.~\ref{fig:stability_josephson} a) for 
$\epsilon \gtrsim 1.325$. Fig.~\ref{fig:mqst} depicts the 
one- and two-dimensional absorption images in $x$-direction 
of this state, where $|\Psi_{y,z}|^2$ is calculated analogously 
to \eqref{psi_square}. This state is robust with respect
to small experimental imperfections like small perturbations in the 
horizontality of the toroidal trap. Indeed, the whole trap, as well as 
exclusively, the toroidal trap, may be rotated by several degrees without 
destroying this MQST-like state. Further imperfections like 
gradients in the magnetic field may affect in which well the 
particles of the condensate accumulate, but we do not expect any
further consequences due to the small magnitude of the gradient 
(usually $\approx$ 1 Gauss/cm). Nonetheless, the effect of small gradients 
upon the accumulation of the particles can be investigated by
aligning the dipoles parallel to the $y$-axis. 
Note that disturbances in general can result in a shift of the 
stability diagram, thus slightly changing the parameters where 
structured states are expected to appear.

Signatures of this MQST-like state can also be 
observed in the dynamics of Josephson junctions, where a 
large initial imbalance in the occupation of the wells cannot 
be periodically compensated and is therefore dominated by 
the self trapping in one well. This results in a running 
phase mode, where the creation of vortex-antivortex pairs 
in regions of small density is responsible for a 
phase shift \cite{Abad2011_2}.   

\section{Conclusion}
\label{sec:conclusion}

We have presented a series of stability and phase diagrams 
with respect to parameters which are of great importance 
for the experimental investigation of a dipolar BEC. Our 
results are valid for all dipolar systems, but we have 
adjusted all parameters to fit the needs of a condensate 
with $^{164}$Dy. We studied the expansion dynamics of a 
dipolar condensate and found two sorts of dipolar signatures 
occurring during TOF sequences: The effects of the DDI are 
noticeable in that the width of the spatially expanding 
cloud is not only governed by the anisotropic momentum 
distribution but also by the direction of the polarization 
axis. In addition, structured states occur during the TOF. 
The analysis of stability diagrams shows that 
structured ground states can be found along the stability 
border and that their visibility is large enough for modern 
imaging systems. We have shown that the realization of a 
self-induced Josephson junction with $^{164}$Dy is restricted 
to a rather small amount of particles. However, this can be 
compensated by a rotation of the external confinement or 
the polarization axis. Finally, the analysis of the stability 
diagram shows that large values of $\epsilon$ allow for a 
macroscopic quantum self-trapping state. Further investigations 
should include the determination of the scattering length 
$a$ of $^{164}$Dy, which can be accomplished by 
a comparison of experimentally measured and theoretically 
calculated in-trap radii and stability diagrams. Our results 
should therefore stimulate experimental 
efforts to study dipolar BECs.      

\section{Acknowledgements}

We thank Holger Kadau, Thomas Maier, Matthias Schmitt 
and Tilman Pfau for valuable discussions.


%

\end{document}